\documentclass[aps,11pt]{revtex4}
\usepackage{dcolumn}
\usepackage{graphicx}
\usepackage{amsmath}
\usepackage{amsfonts}
\usepackage{amssymb}
\usepackage{psfrag}
\usepackage{wrapfig}
\usepackage{subfigure}
\usepackage{makeidx}
\usepackage{bm}
\usepackage{epsf}
\usepackage{hyperref}
\usepackage{color}
\usepackage{cases}
\makeatletter

\newcommand{\Rmnum}[1]{\expandafter\@slowromancap\romannumeral #1@}

\makeatother

\begin{document}
\title{$P-V$ criticality and Joule-Thomson Expansion of Hayward-AdS black holes in 4D Einstein-Gauss-Bonnet gravity}

\author{Ming Zhang$^{1}$\footnote{e-mail: mingzhang0807@126.com; zhangming@xaau.edu.cn}, Chao-Ming Zhang$^{2}$\footnote{e-mail: 843395448@qq.com}, De-Cheng Zou$^{2}$\footnote{e-mail: dczou@yzu.edu.cn} and Rui-Hong Yue$^{2}$\footnote{e-mail: rhyue@yzu.edu.cn}}

\address{$^{1}$Faculty of Science, Xi'an Aeronautical University, Xi'an 710077 China\\
$^{2}$Center for Gravitation and Cosmology, College of Physical Science and Technology, Yangzhou University, Yangzhou 225009, China}

\date{\today}

\begin{abstract}
\indent

In this paper, the $P-V$ criticality and Joule-Thomson Expansion of Hayward-AdS black holes in 4D Einstein-Gauss-Bonnet gravity are studied in the extended phase space. We find the black hole always exhibits a phase transition similar to that of the Van der Waals system for any arbitrary positive parameters $\alpha$ and $g$. We also study the dependence of $\alpha$ and $g$ on the inversion curves and plot the inversion and isenthalpic curves in the $T-P$ plane, which can determine the cooling-heating regions.
\end{abstract}

%\keywords{}

\maketitle

\section{Introduction}
\label{intro}

Following the advent of string theory, extra dimensions were promoted from an
interesting curiosity to a theoretical necessity since superstring theory requires an eleven-dimensional spacetime to be consistent from a quantum point of view (\cite{Horava:1996ma}-\cite{Randall:1999vf}). Among the higher curvature gravities, the most extensively studied theory is the so-called Gauss-Bonnet gravity. However, the GB term's variation is a total derivative in 4D, which has no contribution to the gravitational dynamics. Therefore, for non-trivial gravitational dynamics one requires $D\geq 5$. Recently, Glavan and Lin \cite{Glavan:2019inb} suggested a novel theory of gravity in 4-dimensional spacetime which called ``4D Einstein Gauss-Bonnet gravity''(EGB). By rescaling the GB coupling constant $\alpha \to \alpha/(D-4)$ and defining the 4-dimensional theory as the limit $D\to 4$, the Gauss-Bonnet term could give rise to non-trivial dynamics. Interestingly, the solution of the same form has been presented in the conformal anomaly inspired gravity\cite{Cai:2009ua,Cai:2014jea}.
Furthermore, the spherically symmetric black hole solutions have been also constructed in this paper. The generalization to other black holes has also appeared, for instance the charged AdS case\cite{Fernandes:2020rpa}, Lovelock\cite{Konoplya:2020qqh,Casalino:2020kbt}, rotating\cite{Kumar:2020owy,Ghosh:2020vpc}, Born-Infeld\cite{Yang:2020jno}, Bardeen\cite{Kumar:2020uyz}  Hayward\cite{Kumar:2020xvu}, etc. There also some important properties of the related 4D EGB black holes have been studied, such as the spinning test particle in the black hole\cite{Zhang:2020qew}, the causality\cite{Ge:2020tid}, the stability and shadows\cite{Guo:2020zmf,Wei:2020ght}.

In the black hole physics, the thermodynamical phase transition of black hole is always a hot topic. Recently, the thermodynamics of AdS black holes has been investigated in the extended phase space, where the cosmological constant is treated as the pressure of the system \cite{Kastor:2009wy,Kubiznak:2012wp}.
It was found that a first order small and large black holes phase transition is allowed and the $P-V$ isotherms are analogous to the Van der Waals fluid.
More discussions including reentrant phase transitions and more general Van der Waals behavior in this direction can be found as well \cite{Gunasekaran:2012dq,Hendi:2012um,Zhao:2013oza,Zou:2013owa,
Dehghani:2014caa,Hennigar:2015esa,Zhang:2014jfa,Altamirano:2014tva,
Wei:2015iwa,Xu:2014kwa,Sadeghi:2016dvc,Hansen:2016ayo}.
In addition, there is a well-known process in classical thermodynamics, called Joule-Thomson expansion, was generalized to charged AdS black holes in Ref.\cite{Okcu:2016tgt}. The inversion and isenthalpic curves were obtained and the heating-cooling regions were illustrated in the $T-P$ plane. Subsequently, Joule-Thomson expansions in various black holes have been studied extensively, such as 4D Gauss-Bonnet AdS black hole\cite{Hegde:2020xlv}, Born-Infeld AdS black hole\cite{Bi:2020vcg}, charged AdS black hole in massive gravity\cite{Nam:2020gud}, Lovelock AdS black hole\cite{Mo:2018qkt}, 5D Einstein-Maxwell-Gauss-Bonnet-AdS black hole\cite{Haldar:2018cks}, hyperscaling violating black hole\cite{Sadeghi:2020bon}, charged AdS black holes in
the Rastall gravity\cite{Meng:2020csd}, Bardeen-AdS black hole\cite{Singh:2020xju,Li:2019jcd}, Hayward-AdS black hole\cite{Guo:2019gkr} and so on.

On the other hand, the regular black holes\cite{Bardeen:1968,Debnath:2015hea,Pourhassan:2016qoz,DeLorenzo:2014pta,Kumar:2020bqf} have attracted much attention recently, which could provide a new window of physics to understand the nature of black hole singularities. The thermodynamic properties of these regular black holes have been investigated in Refs.\cite{Flachi:2012nv,Abchouyeh:2013qca,Hayward:2005gi,Halilsoy:2013iza}, especially many interesting properties in Hayward-AdS black holes\cite{Guo:2019gkr,Abbas:2014oua,Rodrigue:2018lzp,Contreras:2018gpl}. In this paper, we investigate the $P-V$ criticality and Joule-Thomson expansion of 4D Hayward-AdS EGB black hole in the extended phase space.

The paper is organized as follows: in Sect.\ref{pv}, we study the thermodynamic and $P-V$ criticality of the Hayward-AdS black holes in 4-dimensional EGB gravity in the extended phase space. Then, in Sect.\ref{2s} we give discussions for the Joule-Thomson expansion of the Hayward-AdS black holes in 4-dimensional EGB gravity, which include the Joule-Thomson coefficient, the inversion curves, the minimum inversion temperature and the isenthalpic curves. Furthermore, we discussed the influence of the GB coefficient and the charge $g$ on the inversion curves. We end the paper with closing remakes in the last section.

\section{Thermodynamics and phase transition of the black hole}
\label{pv}

The action of $D$-dimensional fully interacting theory of gravity minimally coupled to nonlinear electrodynamics (NED) in the presence of a negative cosmological constant $\Lambda\equiv -\frac{(D-1)(D-2)}{2l^2}$ is given by
\begin{eqnarray}
{\cal S}&=&\frac{1}{16\pi}\int d^{D}x \sqrt{-g}\left[R+\frac{(D-1)(D-2)}{l^2}+\frac{\alpha}{D-4}{\cal G}+{\cal L}(F) \right], \label{action}
\end{eqnarray}
where the Gauss-Bonnet term is ${\cal G}=R^2-4R_{\mu\nu}R^{\mu\nu}+R_{\mu\nu\rho\sigma}R^{\mu\nu\rho\sigma}$, the Gauss-Bonnet coefficient $\alpha$ with dimension $(length)^2$ is positive in the heterotic string theory. $F=F_{\mu\nu}F^{\mu\nu}/4$ where $F_{\mu\nu}=\partial_\mu A_\nu-\partial_\nu A_\mu$ is a field strength tensor tensor. $A_{\mu}$ is the gauge potential with corresponding tensor field ${\cal L}(F)$.

Here we consider the following $D$ dimensional Lagrangian density of NED field\cite{AyonBeato:2000zs,Fernando:2016ksb}
\begin{eqnarray}
  {\cal L}(F)=\frac{(D-1)(D-2)M}{4g^2}\frac{(2g^2F)^\frac{D-1}{D-2}}{(1+(\sqrt{2g^2F})^{\frac{D-1}{D-2}})^2}
\end{eqnarray}
where $g$ is the magnetic monopole charge and
\begin{equation}
F=\frac{g^{2(D-3)}}{2r^{2(D-2)}}.
\end{equation}

Taking the limit $D\to 4$ \cite{Glavan:2019inb}, four dimensional static and spherically symmetric Hayward black hole solution in EGB gravity are obtained as
\begin{eqnarray}
&&ds^2=-f(r)dt^2+\frac{1}{f(r)}dr^2+r^2d\Omega_{D-2},\label{metric}\\
&&f(r)=1+\frac{r^2}{2\alpha}\left(1-\sqrt{1+4\alpha\left(\frac{2M}{r^3+g^3}-\frac{1}{l^2}\right)} \right),\label{solution}
\end{eqnarray}
where $M$ is the ADM mass of the black hole.
In the extended phase space, the cosmological constant $\Lambda$ is regarded as a variable and also identified with the thermodynamic pressure $P=-\frac{\Lambda}{8\pi}=\frac{3}{8\pi l^2}$ in the geometric units $G_N=\hbar=c=k=1$. In the low energy effective action of heterotic string theory, $\alpha$ is proportional to the inverse string tension with positive coefficient, thus we will only consider the positive GB coefficient $\alpha$ in the following discussion. If we take $g\to 0$, the solution $f(r)$ reduces to 4D EGB AdS case.

In terms of the horizon radius $r_+$, the mass $M$ and Hawking temperature $T$ of
4D Hayward-AdS EGB black holes can be written as
\begin{eqnarray}
&&M=\frac{(g^3+r_+^3)\left[r_+^4+l^2(r_+^2+\alpha) \right]}{2l^2 r_+^4},\label{M}\\
&&T=\frac{f'(r)}{4\pi}|_{r=r_+}=\frac{8\pi r_+^7 P+r_+^3(r_+^2-\alpha)-2g^3(r_+^2+2\alpha)}{4\pi r_+(g^3+r_+^3)(r_+^2+2\alpha)} \label{TS}
\end{eqnarray}

It's worth noticing that the Hayward black hole belongs to the non-linear electrodynamics black hole solutions, to be more precisely those in which the matter Lagrangian depends on the black hole mass\cite{Ma:2014qma}. For this class of black holes, in order to obey the Wald's formula\cite{Wald:1993nt} and Visser's result\cite{Visser:1993qa}, some authors\cite{Ma:2014qma,Balart:2017dzt,Zhang:2016ilt,Gulin:2017ycu} suggested that the first law of black hole thermodynamics need to be modified. Therefore, the general form of the first law applied to Hayward-AdS black hole can be written as
\begin{eqnarray}
  (1-\phi_M)dM=TdS+VdP+\phi_g dg+\phi_{\alpha}d\alpha \label{eq_first}
\end{eqnarray}
where the pressure $P=-\frac{\Lambda}{8\pi}$, $V$ and $\phi_{\alpha}$ are the conjugate potentials of the pressure and Gauss-Bonnet coupling respectively. The thermodynamics variables appearing in Eq.\ref{eq_first} are given by
\begin{eqnarray}
  &&T=\frac{3r_+^7-2g^3 l^2(r_+^2+2\alpha)+l^2(r_+^5-r_+^3\alpha)}{4\pi l^2r_+(r_+^2+2\alpha)(r_+^3+g^3)}; \label{tem}\\
  &&P=\frac{3}{8\pi l^2};~~V=\frac{4\pi r_+^3}{3};~~ S=\pi r_+^2+4\pi\alpha\ln{r_+}; \label{pvs}\\
  &&\phi_M=1-\frac{r_+^3}{r_+^3+g^3};~~ \phi_g=\frac{3g^2\left[r_+^4+l^2(r_+^2+\alpha) \right]}{2l^2 r_+(r_+^3+g^3)};\\
  &&\phi_{\alpha}=\frac{1}{2r_+}-4\pi T\ln{r_+}
\end{eqnarray}

From the Hawking temperature (\ref{TS}), we can obtain the equation of state
\begin{eqnarray}
P=\frac{2g^3(r_+^2+2\alpha)-r_+^3(r_+^2-\alpha)}{8\pi r_+^7}+\frac{(g^3+r_+^3)(r_+^2+2\alpha)}{2r_+^6}T \label{eos}
\end{eqnarray}
As usual, a critical point occurs when $P$ has an inflection point,
\begin{eqnarray}
\frac{\partial P}{\partial r_+}\Big|_{T=T_c, r_+=r_c}
=\frac{\partial^2 P}{\partial r_+^2}\Big|_{T=T_c, r_+=r_c}=0.\label{inflection}
\end{eqnarray}
Then we can obtain the equation for the critical horizon radius
\begin{eqnarray}
&&r_c^{10}-12\alpha r_c^8-28g^3 r_c^7-12\alpha^2 r_c^6-192\alpha g^3 r_c^5-20g^6 r_c^4 \nonumber\\
&&-288\alpha^2 g^3 r_c^3-108\alpha g^6 r_c^2-168\alpha^2 g^6=0. \label{crirc}
\end{eqnarray}
With Eq.\ref{eos}, Eq.\ref{inflection} and Eq.\ref{crirc}, the critical temperature and critical pressure can be written as
\begin{eqnarray}
&&T_c=\frac{2\alpha r_c^6(5r_c^2+6\alpha)+g^3(23r_c^7+178\alpha r_c^5+288\alpha^2 r_c^3)+4g^6(5r_c^4+27\alpha r_c^2+42\alpha^2)}{2\pi r_c^6[r_c^5+6\alpha r_c^3+4g^3(r_c^2+3\alpha)]} \\
&&P_c=\frac{\alpha r_c^6(7r_c^2+10\alpha)+2g^3(9r_c^7+76\alpha r_c^5+130\alpha^2 r_c^3)+2g^6(9r_c^4+50\alpha r_c^2+80\alpha^2)}{8\pi r_c^7(r_c^5+6\alpha r_c^3+4g^3(r_c^2+3\alpha))}
\end{eqnarray}
Here $r_c$, $T_c$ and $P_c$ are all positive cause of the critical point to be physical.

Now we consider the critical behaviors of 4D Hayward-AdS EGB black hole in the extended phase space. For Eq.\ref{crirc}, we can see the equation of critical horizon radius contains the higher-order polynomials, which means the analytic solution is not possible. However, we care more about the number of the physical points which determine the type of phase transition the system contains. The number of the positive roots for a higher order equation can be distinguished by the Descartes' rule of signs\cite{Descartes}, which is expressed as : ``An equation can have as many positive roots as it contains changes of sign, from $+$ to $-$ or from $-$ to $+$.'' By using this rule, we can immediately indicate that there is one and only one positive root of the Eq.\ref{crirc} for arbitrary positive $\alpha$ and positive $g$. From Eq.\ref{eos} and Eq.\ref{inflection}, we can also distinguish the critical temperature $T_c$ and critical pressure $P_c$ are always positive with a positive critical radius $r_c$, which means the system always has one physical critical point corresponding to a Van der Waals like phase transition for arbitrary positive $\alpha$ and positive $g$.
For instance, we can obtain a critical point with $r_c=3.252$, $T_c=0.03461$ and $P_c=0.002076$ by fixed $\alpha=0.1$ and $g=1$.

Moreover, in the case of $g=0$ or $\alpha=0$, the critical points can be analytically solved from Eq.\ref{eos}, Eq.\ref{inflection} and Eq.\ref{crirc},
\begin{subequations}
  \begin{numcases}{}
   g=0;~~ r_c=\sqrt{2}\sqrt{3\alpha+2\sqrt{3}\alpha}\\
   T_c=\frac{1+\sqrt{3}}{2(3+\sqrt{3})\sqrt{6+4\sqrt{3}}\pi\sqrt{\alpha}}, \\
   P_c=\frac{13+7\sqrt{3}}{3168\pi\alpha+1824\sqrt{3}\pi\alpha},
  \end{numcases}
\end{subequations}
and
\begin{subequations}
  \begin{numcases}{}
   \alpha=0;~~ r_c=[2(7+3\sqrt{6})]^{1/3}g \\
   T_c=\frac{3+2\sqrt{6}}{4\pi(3+\sqrt{6})(14+6\sqrt{6})^{1/3}g}, \\
   P_c=\frac{9(5+2\sqrt{6})}{16\times 2^{2/3}\pi(3+\sqrt{6})(7+3\sqrt{6})^{5/3}g^2},
  \end{numcases}
\end{subequations}
which covered the results of 4D EGB AdS black hole\cite{Zhang:2020khz} and 4D Hayward AdS black hole\cite{Kumara:2020mvo}.

The behavior of Gibbs free energy $G$ is important to determine the thermodynamic phase transition. The free energy $G$ obeys the following thermodynamic relation $G=M-TS$ with
\begin{eqnarray}
G&=&\left[\frac{4}{3}\pi(r_+^3+g^3)-\frac{2\pi r_+^6(r_+^2+4\alpha\ln{r_+})}{(r_+^3+g^3)(r_+^2+2\alpha)}\right]P
+\frac{(r_+^3+g^3)(r_+^2+\alpha)}{2r_+^4} \nonumber\\
&+&\frac{[-r_+^5+r_+^3\alpha+2g^3(r_+^2+2\alpha)](r_+^2+4\alpha\ln{r_+})}
{4r_+(r_+^3+g^3)(r_+^2+2\alpha)}.\label{free}
\end{eqnarray}
Here $r_+$ is understood as a function of pressure and temperature, $r_+=r_+(P,T)$, via equation of state (\ref{eos}).

\begin{figure}[htb]
\centering
\subfigure[$T_c=0.03461$]{\label{fig:subfig:pr} %% label for first subfigure
\includegraphics[width=2.8in]{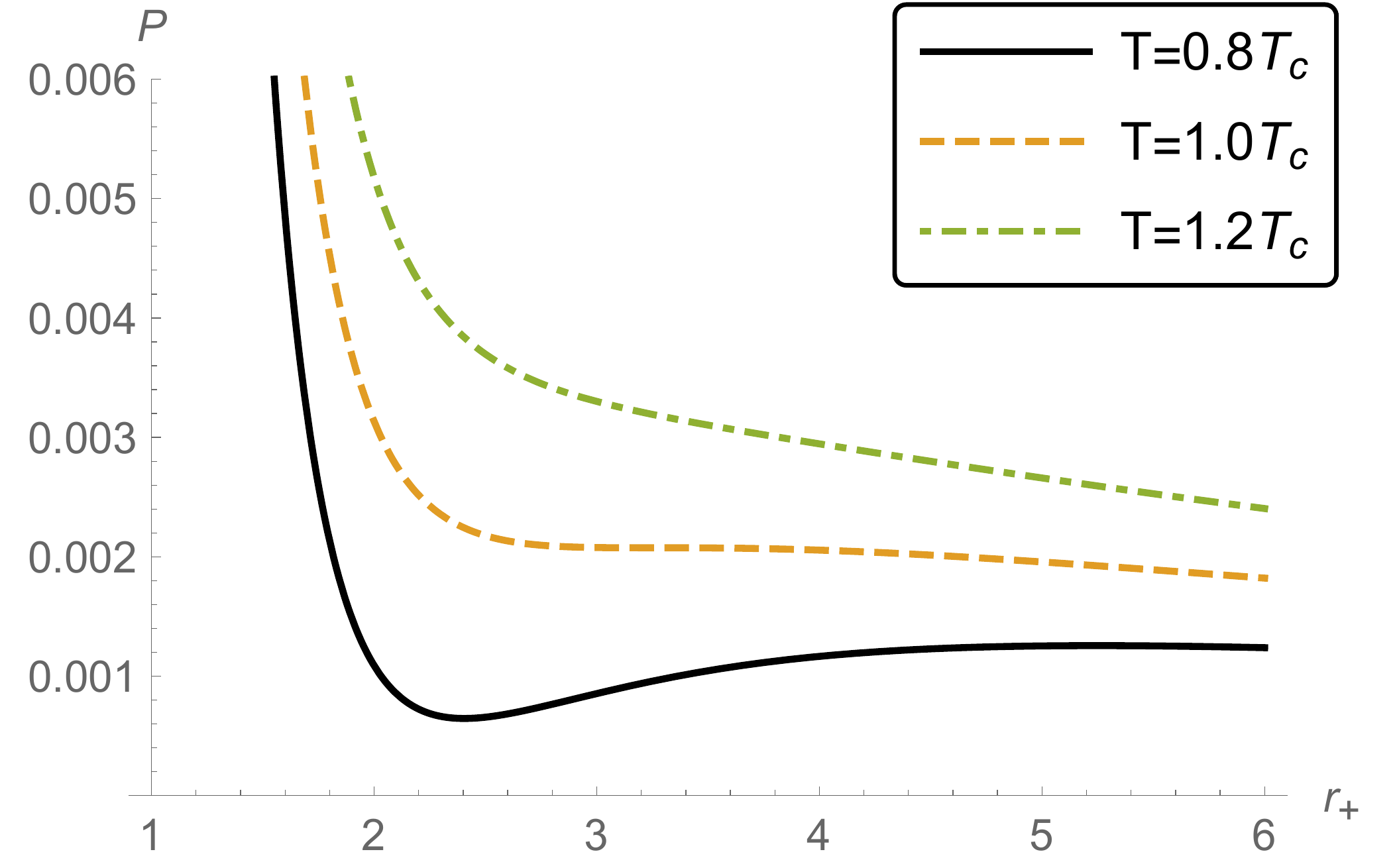}}
\hfill
\subfigure[$P_c=0.002076$]{\label{fig:subfig:gt} %% label for first subfigure
\includegraphics[width=2.8in]{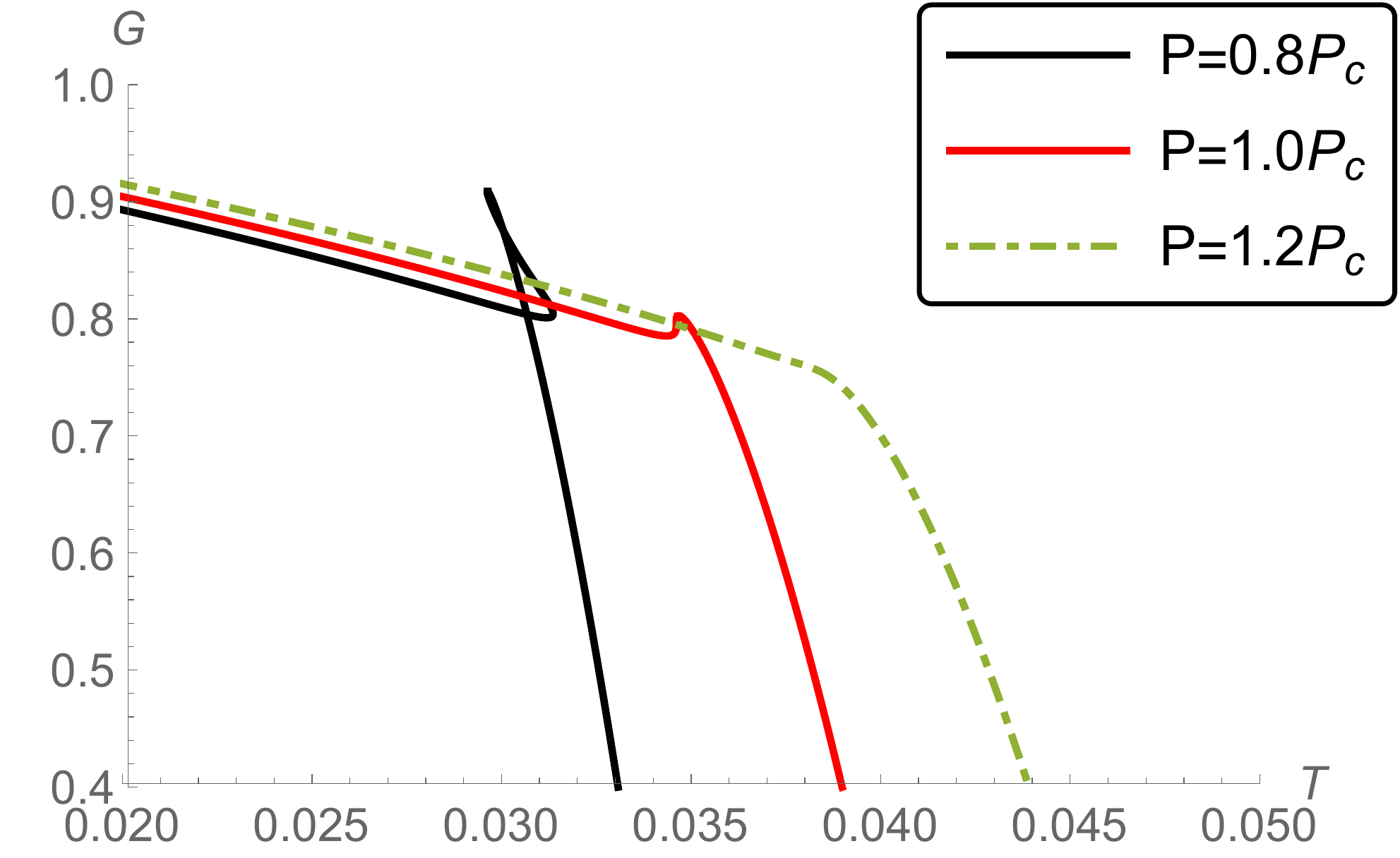}}
\hfill
\caption{The $P-r_+$ and $G-T$ diagram of Hayward AdS black holes with $\alpha=0.1, g=1$.}\label{figpg}
\end{figure}

In Fig.\ref{fig:subfig:pr}, we plot the $P-r_+$ isotherm diagram around the critical temperature $T_c$ for the 4D Hayward-AdS EGB black holes. The dotted line with $T>T_c$ corresponds to the ``idea gas'' phase behavior, and when $T<T_c$ the Van der Waals like small/large black hole phase transition will appear. Fig.\ref{fig:subfig:gt} depicts that the Gibbs free energy as a function of black hole temperature for three different values of pressure. It demonstrates a ``swallow tail'' behavior below the critical pressure, which means the system contains a Van der Waals like first order phase transition.

\section{Joule-Thomson expansion of the black hole}
\label{2s}

During the throttling process, heating or cooling is an interesting feature in a Van der Waals system. In the above section, we find that the phase structure of the 4D Hayward-AdS EGB black hole system can be analogous to that of the Van der Waals system. Therefore, we investigate Joule-Thomson expansion for the black hole in this section. It is already known that the AdS black holes exhibit the throttling process\cite{Okcu:2016tgt,Okcu:2017qgo,Rizwan:2018mpy}. During this expansion process, the enthalpy remains constant, and the black hole mass is considered as the enthalpy in the AdS space. The Joule-Thomson coefficient $\mu$ is defined as
\begin{eqnarray}
  \mu=(\frac{\partial{T}}{\partial{P}})_H=(\frac{\partial{T}}{\partial{P}})_M
\end{eqnarray}
This coefficient characterizes the expansion and plays an important role as its sign describes whether the heat is absorbed or evolved during the expansion process. It is easy to see that the system will experience a cooling (heating) process with $\mu>0$ ($\mu<0$), caused by the change in pressure is always negative during expansion. By passing the inversion point ($[T_i, P_i]$ outcome of $\mu=0$), this process will change to heating (cooling).

In the following, the parameters $\alpha$ and $g$ are all kept fixed. Since the mass and entropy of the black hole are state functions, the differential $dM$ and $dS$ can be expressed as
\begin{eqnarray}
  &&dM=(\frac{\partial M}{\partial T})_{P,\alpha,g}dT+(\frac{\partial M}{\partial P })_{T,\alpha,g}dP \label{eqdh}\\
  &&dS=(\frac{\partial S}{\partial T})_{P,\alpha,g}dT+(\frac{\partial S}{\partial P })_{T,\alpha,g}dP \label{eqds}
\end{eqnarray}
By substituting Eq.\ref{eqds} to the general form of the first law Eq.\ref{eq_first}, we can get
\begin{eqnarray}
  (1-\phi_M)dM&=&T(\frac{\partial S}{\partial T})_{P,\alpha,g}dT+\left[T(\frac{\partial S}{\partial P})_{T,\alpha,g}+V\right]dP \label{eqdhh}
\end{eqnarray}
When the black hole system goes through a isenthalpy process $dM=0$, from Eq.\ref{eqdh} and Eq.\ref{eqdhh}, one can obtain
\begin{eqnarray}
  \mu&=&(\frac{\partial{T}}{\partial{P}})_M=-\frac{(\partial M/\partial P)_{T,\alpha,g}}{(\partial M/\partial T)_{P,\alpha,g}} \nonumber\\
  &=&-\frac{T(\frac{\partial S}{\partial P})_{T,\alpha,g}+V}{T(\frac{\partial S}{\partial T})_{P,\alpha,g}} \label{eqmu}
\end{eqnarray}
Note that as mentioned above, for regular black hole the first law need to be modified as Eq.\ref{eq_first} and the Maxwell relation is no longer satisfied. Therefore, the expression $\mu=\frac{1}{C_P}[T(\frac{\partial V}{\partial T})_P-V]$ in the thermodynamics is not valid in the regular black hole.

\begin{figure}[htb]
\centering
\subfigure[$g=1$]{\label{fig:subfig:g1} %% label for first subfigure
\includegraphics[width=2.8in]{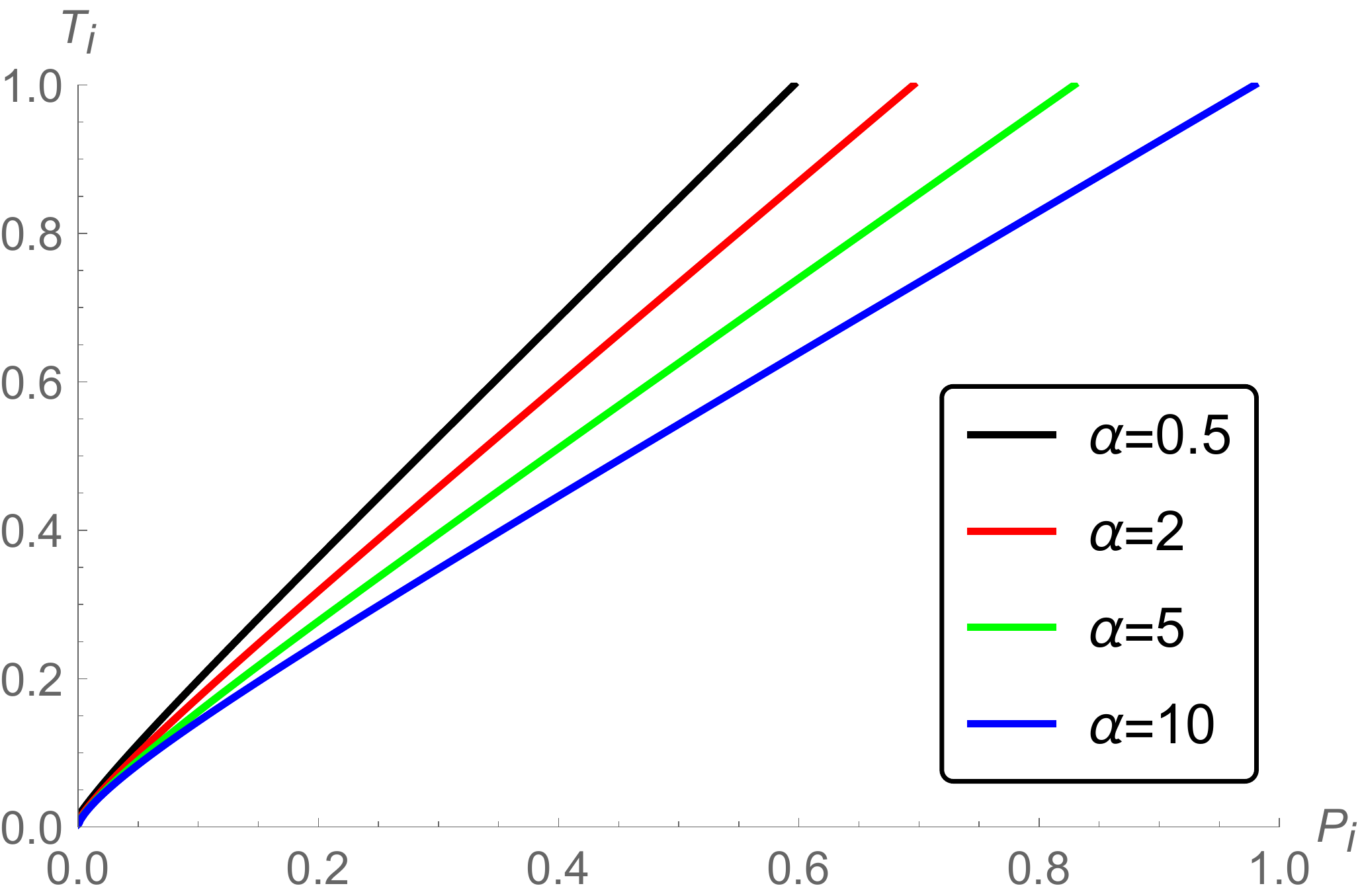}}
\hfill
\subfigure[$\alpha=1$]{\label{fig:subfig:a1} %% label for first subfigure
\includegraphics[width=2.8in]{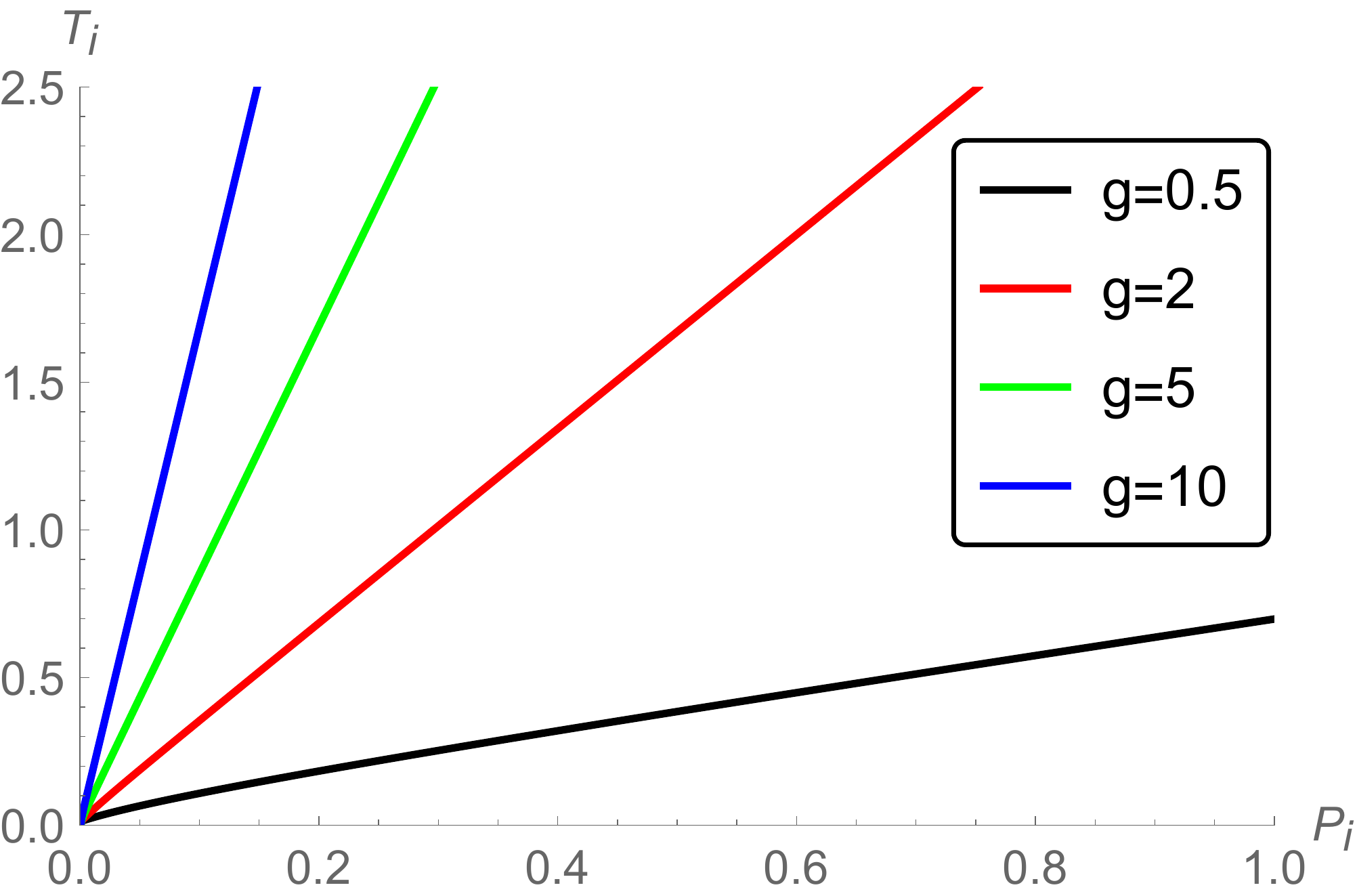}}
\hfill
\caption{Inversion curves for 4D Hayward-AdS EGB black holes in T-P plane. From bottom to top, the left curves correspond to $g=1$ and $\alpha=10,5,2,0.5$, the right curves correspond to $\alpha=1$ and $g=0.5,2,5,10$.}\label{figti}
\end{figure}

We can obtain the coefficient $\mu$ by using the expression of the entropy, temperature and volume in Eq.\ref{tem} and Eq.\ref{pvs}.
\begin{eqnarray}
  &&\mu=4r_+^3\frac{g^6(r_+^2+2\alpha)^2+r_+^6(-2r_+^4-8P\pi r_+^6+r_+^2\alpha+4\alpha^2)+2g^3A}{3(r_+^3+g^3)(r_+^2+2\alpha)^2B} \\
  &&A=8P\pi r_+^9+16r_+^5\alpha+13r_+^3\alpha^2+4r_+^7(1+6P\pi\alpha) \nonumber\\
  &&B=-r_+^5-8P\pi r_+^7+r_+^3\alpha+2g^3(r_+^2+2\alpha) \nonumber
\end{eqnarray}
Applying $\mu=0$, the inversion temperature of the black hole can be written as
\begin{eqnarray}
  T_i=-V\left(\frac{\partial{P}}{\partial{S}}\right)_{T,\alpha,g}
  =-V\left(\frac{\partial P}{\partial r_+}/\frac{\partial S}{\partial r_+} \right)_{T,\alpha,g} \label{ti}
\end{eqnarray}
Combining the expression of $S$ and $P$ in Eq.\ref{pvs} and Eq.\ref{eos}, the inversion temperature Eq.\ref{ti} can be evaluated and expressed as
\begin{eqnarray}
  T_i=\frac{2g^6(r_{+i}^2+2\alpha)^2+r_{+i}^6 C+2g^3r_{+i}^3 D}{12\pi r_{+i}^4(r_{+i}^3+g^3)(r_{+i}^2+2\alpha)^2} \label{ti1}
\end{eqnarray}
where
\begin{eqnarray}
  &&C=8\pi P_ir_{+i}^6+5\alpha r_{+i}^2+2\alpha^2+r_{+i}^4(48\pi\alpha P_i-1) \label{tia} \\
  &&D=16\pi P_ir_{+i}^6+20\alpha r_{+i}^2+14\alpha^2+r_{+i}^4(48\pi\alpha P_i+5) \label{tib}
\end{eqnarray}
and $P_i$ and $r_{+i}$ represent the inversion pressure and the corresponding horizon radius respectively. On the other hand, according to the definition of the temperature Eq.\ref{TS}, we can also write the inversion temperature as
\begin{eqnarray}
  T_i=\frac{8\pi r_{+i}^7 P_i+r_{+i}^3(r_{+i}^2-\alpha)-2g^3(r_{+i}^2+2\alpha)}{4\pi r_{+i}(g^3+r_{+i}^3)(r_{+i}^2+2\alpha)} \label{ti2}
\end{eqnarray}
Substitute Eqs.\ref{ti1}, \ref{tia} and \ref{tib} to Eq.\ref{ti2}, we can show the inversion temprature and the inversion pressure in terms of the corresponding horizon radius $r_{+i}$,
\begin{eqnarray}
  &&T_i=\frac{-r_{+i}^5+2\alpha r_{+i}^3+g^3(5r_{+i}^2+14\alpha)}{4\pi r_{+i}[r_{+i}^5-2g^3(r_{+i}^2+3\alpha)]} \label{tir}\\
  &&P_i=\frac{-2r_{+i}^{10}+\alpha r_{+i}^8+4\alpha^2r_{+i}^6+g^6(r_{+i}^2+2\alpha)^2+g^3r_{+i}^3(8r_{+i}^4+32\alpha r_{+i}^2+26\alpha^2)}{8\pi r_{+i}^7[r_{+i}^5-2g^3(r_{+i}^2+3\alpha)]} \label{pir}
\end{eqnarray}
Via Eq.\ref{tir} and Eq.\ref{pir}, the inversion curves for different values of $\alpha$ and charge $g$ are plotted in Fig.\ref{figti}. The inversion temperature increases monotonously with the inversion pressure. For the charge $g=1$, Fig.\ref{fig:subfig:g1} exhibits the effect of $\alpha$ on the inversion curves. We can find that with the increasing of $\alpha$, the inversion temperature for given pressure tends to decrease. By fixing $\alpha=1$, Fig.\ref{fig:subfig:a1} shows the effect of charge $g$ on the inversion curves. The inversion temperature increases with the increasing of the charge $g$, which is qualitatively similar to the RN-AdS black holes\cite{Okcu:2016tgt}. Comparing with the Van der Waals fluids, the inversion curve of 4D Hayward-AdS EGB black hole is not closed, which means the black holes always cool above the inversion curve during the Joule-Thomson expansion.

\begin{figure}[htb]
\centering
\label{fig:subfig:tmin} %% label for first subfigure
\includegraphics[width=2.9in]{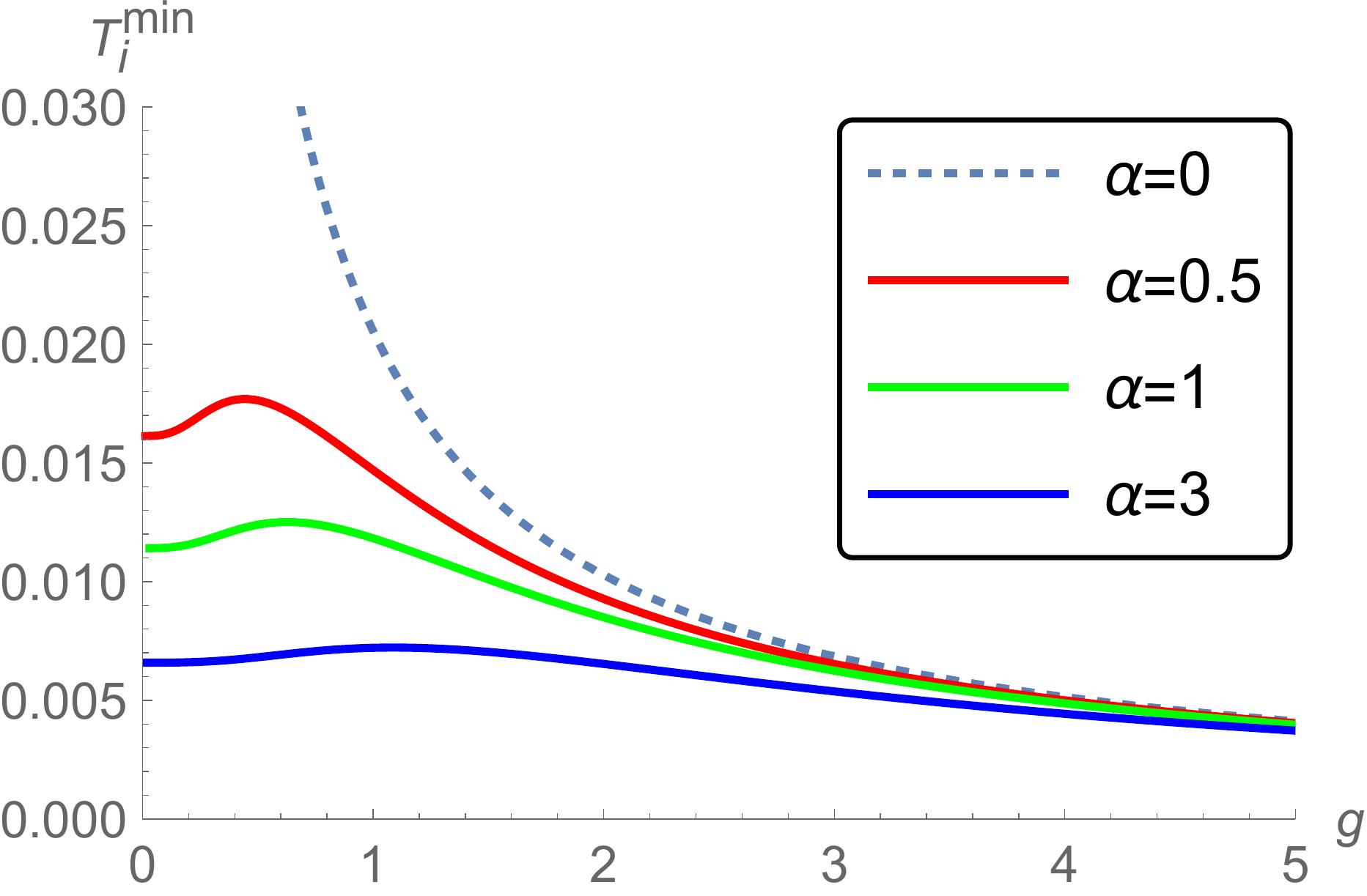}
\caption{The minimum inversion temperature $T_i^{\text{min}}$ versus the charge $g$. From bottom to top, the curves correspond to $\alpha=3,1,0.5,0$.}\label{figtmin}
\end{figure}

The minimum inversion temperature $T_i^{\text{min}}$ occurs at the point $P_i=0$. Since there are higher order terms in $P_i$, the minimum inversion temperature can be obtained numerically. Fig.\ref{figtmin} shows the charge $g$ dependence of $T_i^{\text{min}}$ with different $\alpha$. We can reduce to the case of 4D Hayward-AdS black hole as $\alpha\to 0$.

\begin{figure}[htb]
\centering
\subfigure[$\alpha=1, g=1$]{\label{fig:subfig:a1g1} %% label for first subfigure
\includegraphics[width=2.8in]{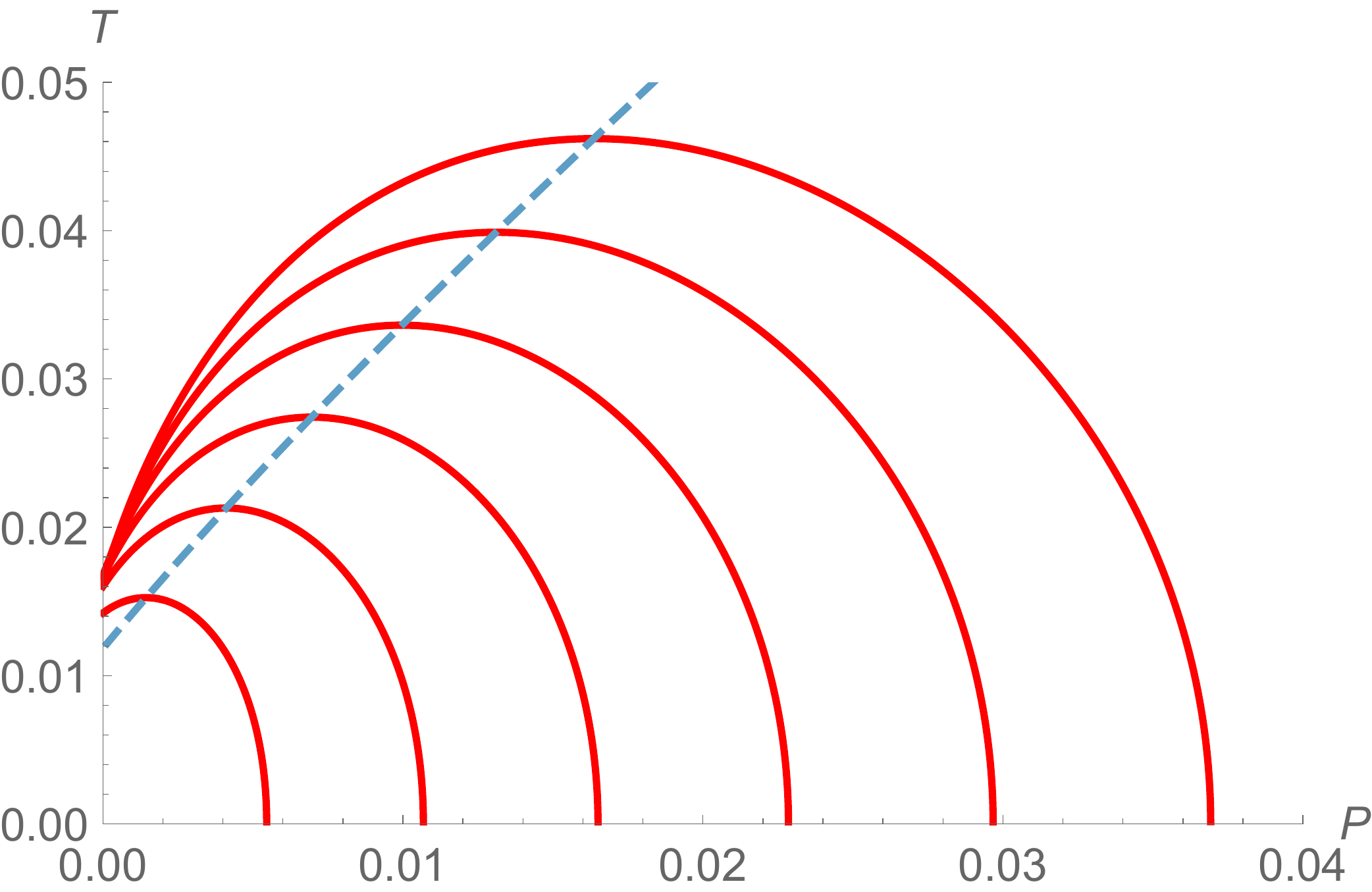}}
\hfill
\subfigure[$\alpha=1, g=2$]{\label{fig:subfig:a1g2} %% label for first subfigure
\includegraphics[width=2.8in]{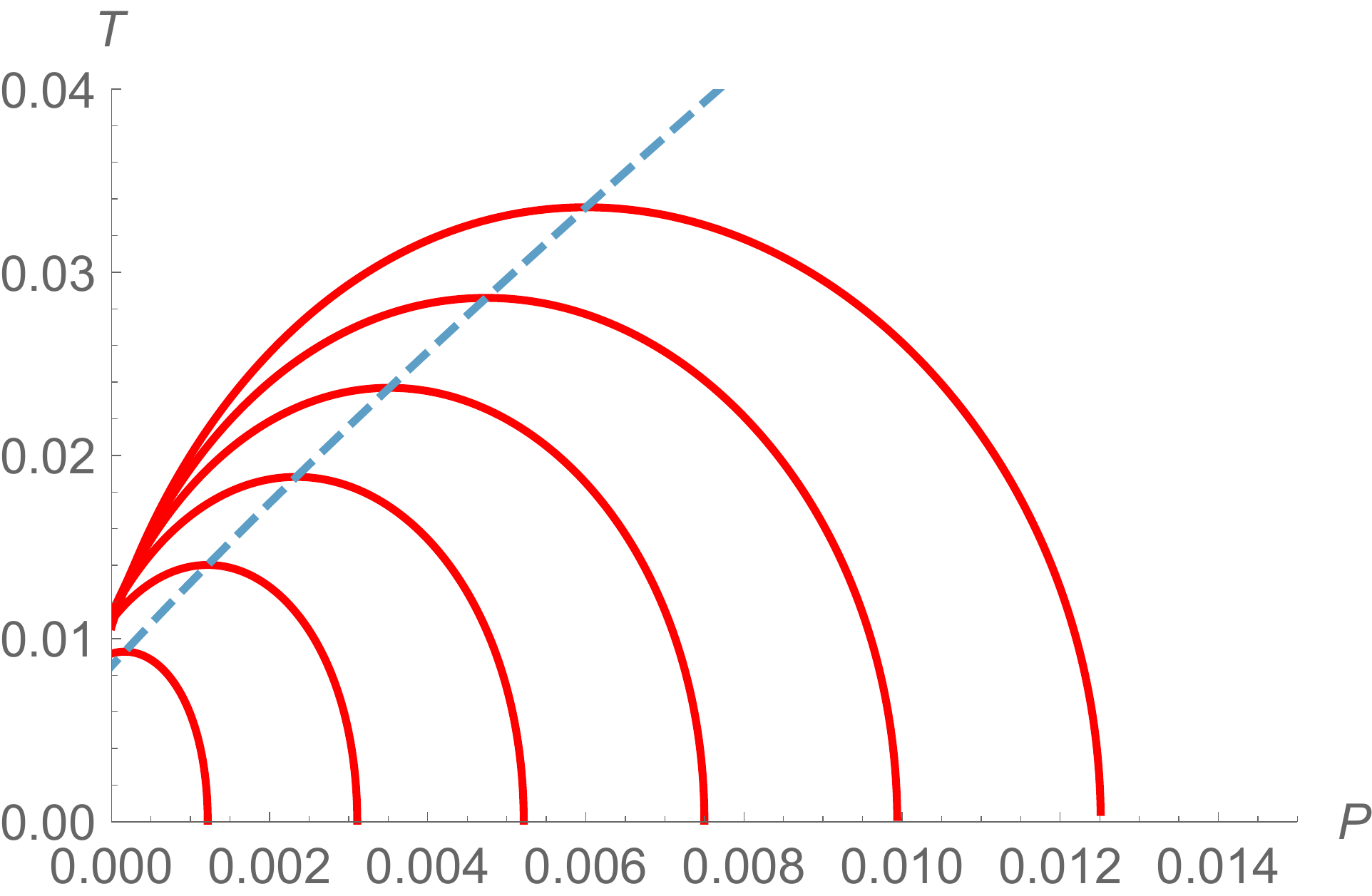}}
\hfill
\subfigure[$\alpha=0, g=1$]{\label{fig:subfig:a0g1} %% label for first subfigure
\includegraphics[width=2.8in]{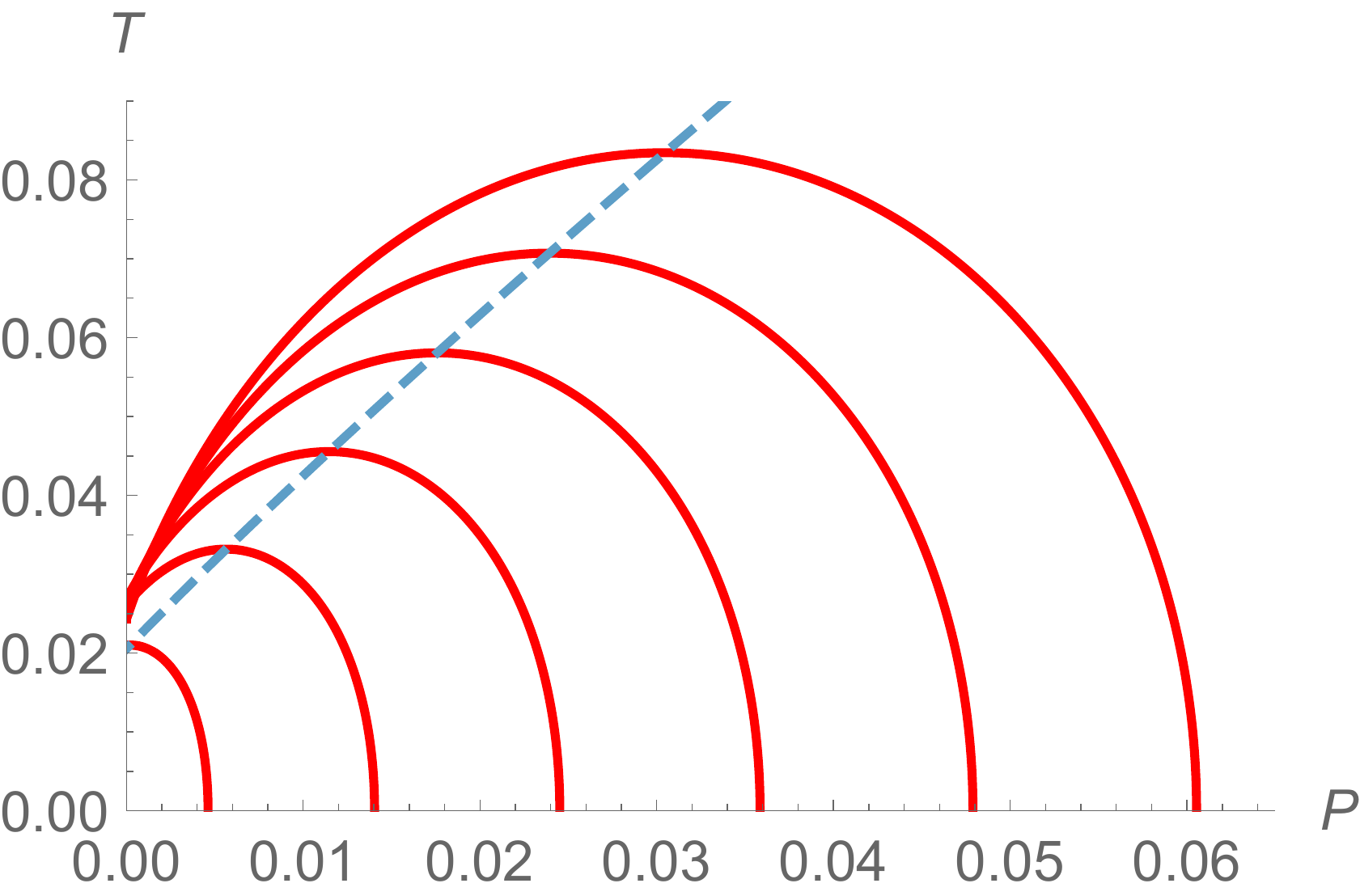}}
\hfill
\subfigure[$\alpha=1, g=0$]{\label{fig:subfig:a1g0} %% label for first subfigure
\includegraphics[width=2.8in]{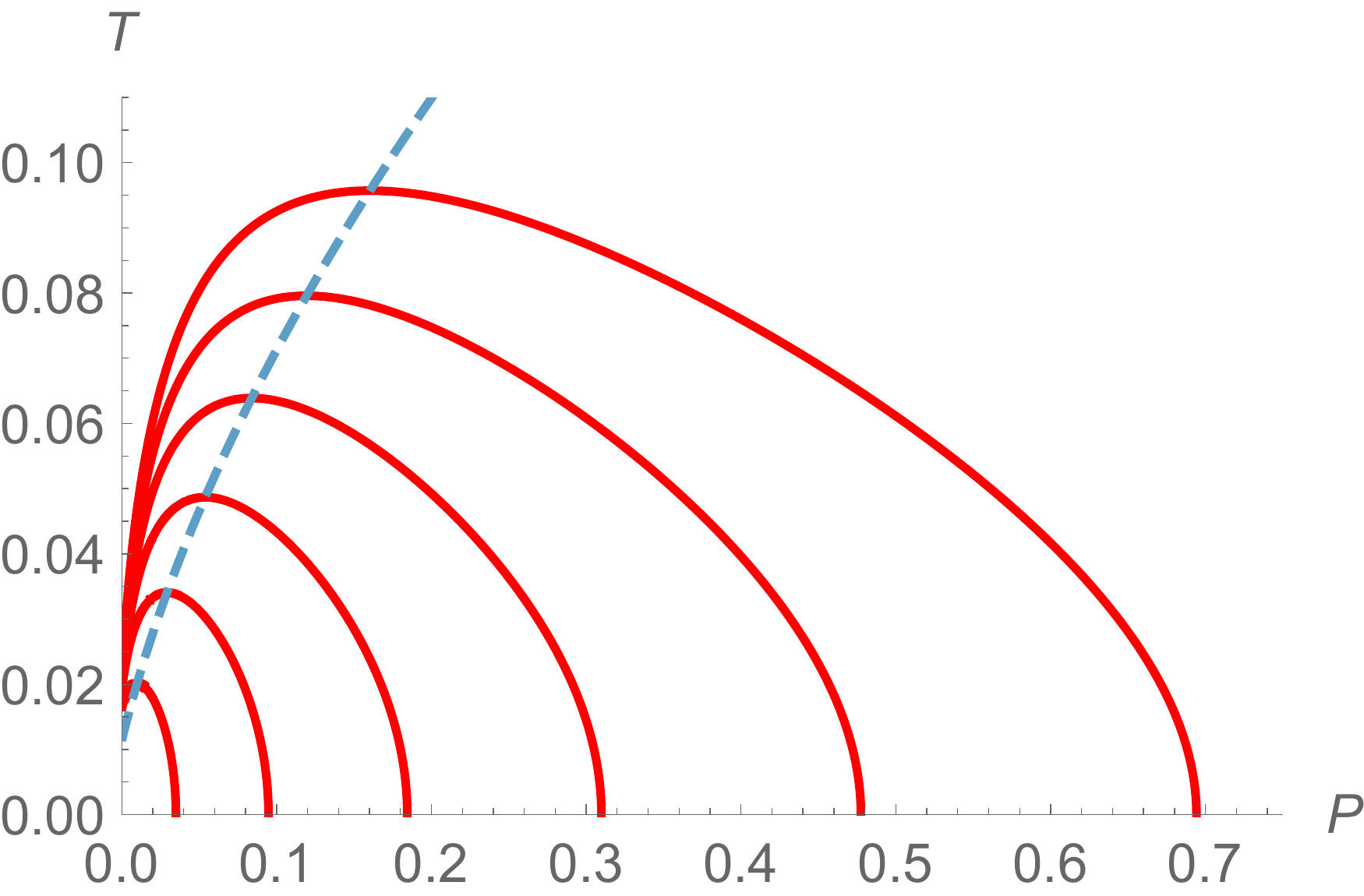}}
\hfill
\caption{Isenthalpic curves for (a) $\alpha=1$, $g=1$, the curves from left to right correspond to $M=1.5,1.6,1.7,1.8,1.9,2.0$ respectively. (b) $\alpha=1$, $g=2$, the curves from left to right correspond to $M=2.3,2.5,2.7,2.9,3.1,3.3$ respectively. (c) $\alpha=0$, $g=1$, the curves from left to right correspond to $M=1.0,1.1,1.2,1.3,1.4,1.5$ respectively. (d) $\alpha=1$, $g=0$, the curves from left to right correspond to $M=1.1,1.2,1.3,1.4,1.5,1.6$ respectively. Note that the inversion curve for $\alpha=1$ is also depicted in both graphs via the dashed line.}\label{fighag}
\end{figure}

In addition, the isenthalpic curves are also of interest considering, since Joule-Thomeson expansion is an isenthalpic process. In the extended phase space, the mass is considered as enthalpy.
In Fig.\ref{fighag}, we plot the isenthalpic curves and the inversion curves in $T-P$ plane by fixing the mass of the black hole. It shows the inversion curve is the dividing line between heating and cooling process. Note that the isoenthalpy curve intersects the inversion curve at the inversion point which also is the maximum point for a specific isenthalpic curve, representing at the inversion point the temperature is highest during the whole Joule-Thomson expansion process.  Above the inversion curve, the slope of the isenthalpic curve is positive, there is a cooling process. On the contrary, the slope changes to negative and the heating occurs below the inversion curve.

\section{Closing remarks}
\label{3s}

In the 4-dimensional Einstein Gauss-Bonnet gravity, we have studied the $P-V$ criticality and Joule-Thomeson expansion of Hayward-AdS black hole in the extended phase space. We obtained the correct thermodynamic variables and the first law, which is contrary to the claims of entropy and volume modification as reported in the literature. We demonstrated the system allows one and only one physical critical point for arbitrary positive parameters $\alpha$ and $g$, which corresponds to the Van der Waals phase transition.

Then, the well-known Joule-Thomeson coefficient $\mu$ is derived and obtained via the first law of black hole thermodynamics. We find that for regular black holes the expression of $\mu$ must be Eq.\ref{eqmu}, since the first law need to be modified and the Maxwell relation is not satisfied in this kind of black hole.
The zero point of $\mu$ is the inversion point which discriminate the cooling process from heating process.

We studied the dependence of $\alpha$ and $g$ on the inversion curves, the results are depicted in Fig.\ref{figti}.  The minimum inversion temperature versus the charge $g$ is displayed on Fig.\ref{figtmin}. We also plot the isenthalpic curves and the inversion curves in Fig.\ref{fighag}, which shows the slope of the inversion curve is always positive. This result means the black hole always cools (heats) above (below) the inversion curve during the expansion. For different values of $\alpha$ and $g$, we can distinguish the cooling and heating regions with the inversion curve.

We would like to thank Xiaoxue Li for useful discussions.

This work is supported by the National Natural Science Foundation of China under Grant Nos.11605152, 11675139, and 51802247.

\end{document}